\documentclass[12pt]{article}
\usepackage{latexsym}

\def\bs{\begin{subequations}}
\def\es{\end{subequations}}

\catcode`\@=11

\newtoks\@stequation
\def\subequations{\refstepcounter{equation}
  \edef\@savedequation{\the\c@equation}%
  \@stequation=\expandafter{\theequation}%   %only want \theequation
  \edef\@savedtheequation{\the\@stequation}% % expanded once
  \edef\oldtheequation{\theequation}%
  \setcounter{equation}{0}%
  \def\theequation{\oldtheequation\alph{equation}}}

\def\endsubequations{\setcounter{equation}{\@savedequation}%
  \@stequation=\expandafter{\@savedtheequation}%
  \edef\theequation{\the\@stequation}\global\@ignoretrue}

\makeatletter%  allow access to internal LaTeX commands
        \renewcommand{\theequation}{\thesection.\arabic{equation}}%
        \@addtoreset{equation}{section}%
\makeatother%  turn off access to internal LaTeX commands

\renewcommand{\thefootnote}{\fnsymbol{footnote}}

\begin{document}

\begin{titlepage}

November 11, 2015 - original submitted to IJMPA 

March 2, 2016 - revised after being accepted by IJMPA

\begin{center}        \hfill   \\
            \hfill     \\
                                \hfill   \\

\vskip .25in

{\large \bf{Toward a Quantum Theory of Tachyon Fields \\}}

\vskip 0.3in

Charles Schwartz\footnote{E-mail: schwartz@physics.berkeley.edu}

\vskip 0.15in

{\em Department of Physics,
     University of California\\
     Berkeley, California 94720}
        
\end{center}

\vfill

\begin{abstract}

We construct momentum space expansions for the wave functions 
that solve the Klein-Gordon and Dirac equations for tachyons, 
recognizing that the mass shell for such fields is very different  
from what we are used to for ordinary (slower than light) particles. 
We find that we can 
postulate commutation or anticommutation rules for the operators that 
lead to physically sensible results: causality, for tachyon fields,
means that there 
is no connection between spacetime points separated by a timelike 
interval. Calculating the conserved charge and 4-momentum for these 
fields allows us to interpret the number operators for particles and 
antiparticles in a consistent manner; and we see that helicity plays 
a critical role for the spinor field. Some questions about Lorentz 
invariance are addressed and some remain unresolved; and we show how to handle the group representation for tachyon spinors.

\end{abstract}

\vfill

\end{titlepage}

\renewcommand{\thefootnote}{\arabic{footnote}}
\setcounter{footnote}{0}
\renewcommand{\thepage}{\arabic{page}}
\setcounter{page}{1}

\section{Introduction}%1%
What of old habits do we keep and what do we change?
That is always the challenging question for theoretical physicists who 
are seeking to  innovate. The idea of tachyons (faster than light particles)  
has been a fascination of some theorists for many decades \cite{ER}; but few 
professional colleagues nowadays grant that idea much credibility.

Among the many objections have been the claims of negative energy 
states, anti-causal behavior, and other bizarrities. My own earlier 
papers have shown that if we work with a conserved 
energy-momentum tensor, we can resolve questions about negative energy 
states \cite{CS1}; and a more recent study showed how quantum wave packet 
considerations eliminated the main complaints about causality 
violation.\cite{CS2} That 
same paper showed the possibly important contributions of (classical) 
tachyons to cosmological studies.

When it comes to quantizing tachyons, especially quantizing a 
tachyonic field theory, the prevalent view \cite{AKS} is that there is 
really nothing that propagates faster than light but it is just some 
unstable state that needs to decay; and everything conforms to ordinary 
concepts of causality.

This paper takes just the opposite approach: assume that tachyons 
might exist as a particle/field system that always propagates faster 
than light. This poses several mathematical challenges, and here we 
show how to manage most of that.  Sections 2 and 3 review basic ideas about the 
mass shell 
and  solutions of the Klein-Gordon equation. Sections 4 and 5 
look at solutions of the Dirac equation, for ordinary particles and for 
tachyons; and we note the suggestion of a critical role for 
helicity  in making a distinction between particles and antiparticles.
Section 6 discusses orthogonality among various plane wave states for 
both equations for both types of particles. In Sections 7 and 8 we 
investigate models of second quantization that may lead to causal 
commutators or anti-commutators. Causality for ordinary (slow) 
particles means that there is no connection between points separated 
by a spacelike interval  
in spacetime, while for tachyons it means just the opposite. 
In Section 9 we identify number operators for particles and 
antiparticles; and in Section 10 we look at 
Lorentz invariance of our results and raise some further 
questions. Section 11 shows how we can handle the group representation problem for a tachyon spinor; and in Section 12 we summarize what has been achieved here and look forward to further work.

\section {Why canonical quantization is wrong here.}%2%

The mathematical procedures known as "canonical quantization" have built into them certain mathematical biases that come from assumed physical behavior of ordinary (slower than light) particles and fields. I believe that some of this is shown, or at least implied, in my 1982 paper; but let me go into this here in some detail. We start with fields defined in 4-dimensional spacetime and we want to ask about how they propagate.

For ordinary particles/fields we assume that some initial solution is contained in some finite volume of 3-space at an initial time $t_0$ and it will also be contained in a finite volume at another time $t$ some finite distance away. This assumption makes sense for any particle/field that can travel no faster than the speed of light; but it is not acceptable for a tachyon, which might travel at arbitrarily high speeds.

For a tachyon field, we look for an alternative geometric arrangement about its propagation that makes sense with the physical understanding that we mean to describe a particle/field that can never propagate slower than the speed of light.  We choose some 2-dimensional surface in 3-dimensional coordinate space, over all values of the time t. Our initial value assumption is that the particle (wave packet) will pass through this surface in some finite time interval; and we can be sure that it will also pass through any other parallel surface, located a finite distance away, in a finite time interval.

Note that this alternative scheme could not be used for a slower-than-light particle/field because there is the possibility for the particle to be at rest, thus it may never pass through the second surface.

(Question: Are we free to use either scheme for light?)

This leaves us with the question, for the tachyon field, whether this chosen surface in 3-space is an open surface or a closed surface. In most of this paper we use an open surface (z= const.); but in the Appendix we chose a closed surface (r=const). I must admit that I do not have a general answer to this question.

\section{Basic issues}%3% 

First we review some basic properties of solutions of the free Klein- 
Gordon  equations for ordinary particles and for tachyons.  They will have 
the space-time behavior in the form of plane waves.
\begin{equation}
\psi(\textbf{x},t) =e^{-ip_{\mu}x^{\mu}} = e^{i\textbf{p}\cdot\textbf{x} -iE t},
\end{equation}
where, for ordinary particles,
\begin{equation}
p^{\mu}p_{\mu} = E^{2} - \textbf{p} \cdot\textbf{p} = E^{2} - p^{2} = +m^{2}.
\end{equation}
This equation, describes the ``mass-shell'' and we see that it 
consists of a two-sheeted hyperboloid: one with $E \ge m$ and the other 
with $E\le -m$.

Under any proper Lorentz transformation, these positive energy and 
negative energy states will remain as two distinct sets of solutions. 
Later, they will be used to describe particles and anti-particles.

However, when we consider tachyons the mass shell is different:
\begin{equation}
p^{\mu}p_{\mu} = E^{2} - \textbf{p} \cdot\textbf{p} = E^{2} - p^{2} = 
-m^{2},
\end{equation}
and this is just a single surface - a hyperboloid of one sheet - in four dimensional space. Positive 
and negative values of E are not separated.

We now introduce spin.

\section{Dirac wavefunction for an ordinary particle}%4%

For the Dirac equation, we have the familiar free-particle solutions,
\begin{equation}
\psi(x) = e^{-ip_{\mu}x^{\mu}} u_{h}(E,\textbf{p}), 
\;\;\;\;\;u_{h}(E,\textbf{p}) = N \left( \begin{array}{c} 
(E+m)|\textbf{p},h> \\ 
hp |\textbf{p},h>\end{array}\right)  \label{b1}
\end{equation}
where $|\textbf{p},h>$ is a 2-component eigenfunction of the spin in the direction of 
$\textbf{p}$ with eigenvalue $ h = \pm 1$.

With a choice of the normalization constant, $N = 
1/\sqrt{2|E+m|}$, we 
calculate the conserved four-current for this wavefunction as,
\begin{equation}
j^{\mu} = \bar{\psi}\; \gamma^{\mu}\; \psi = sign(E) (E, \textbf{p}),
\end{equation}
where $\bar{\psi} = \psi^{\dagger}\gamma^{0}$.  We similarly 
calculate the components of the conserved Energy-Momentum tensor 
\footnote{Some authors have different ways of writing this tensor for 
the Dirac wavefunction. This form is simple and manifestly 
real and symmetric.} as,
\begin{eqnarray}
T^{\mu \nu} = (i/4)[ \bar{\psi}\; \gamma^{\mu}\;\stackrel 
{\leftrightarrow} {\partial^{\nu}} \psi 
+\bar{\psi}\gamma^{\nu}\;\stackrel{\leftrightarrow} 
{\partial^{\mu}}\psi], \;\;\; \stackrel{\leftrightarrow}{\partial} = 
\stackrel{\rightarrow}{\partial} - \stackrel{\leftarrow}{\partial} \\
T^{00} = sign(E) E^{2}, \;\;\;\;\;T^{03}= T^{30} = sign(E) Ep, 
\;\;\;\;\; T^{33} = sign(E) p^{2},
\end{eqnarray}
where we assume that the momentum is in the 3-direction; and all 
other components of the tensor vanish.

In both of those calculations we see that the factor $sign(E)$ may 
cause some confusion, until we introduce annihilation and creation 
operators, which anti-commute with one another, and let us have 
positive counts of particle or antiparticle number as well as energy.

\section{Dirac wavefunction for a Tachyon}%5%

We can start by replacing the mass $m$ in the Dirac equation by an 
imaginary $im$; and we do the same with the plane wave solutions 
(\ref{b1}).  We also need a different definition for the adjoint 
wavefunction,
\begin{equation}
\bar{\psi} = \psi^{\dagger}\gamma^{0}\;\gamma_{5},
\end{equation}
where $\gamma_{5}$ is the unit Hermitian matrix that anticommutes 
with all the $\gamma^{\mu}$.  Then, with the normalization constant 
$N = 1/\sqrt{2p}$, we calculate:
\begin{equation}
j^{\mu} = -h (E,\textbf{p}), \;\;\;\;\; T^{00} = -h E^{2}, \;\;\;\;\; 
T^{03}=T^{30}=-h Ep, \;\;\;\;\;T^{33} = -h p^{2}.
\end{equation}

Here is the strong suggestion that the helicity of the tachyon 
wavefunction  
can serve to distinguish between particle and anti-particle

\section{Orthogonalities}%6%

Next we want to consider the general superposition of plane waves, 
for scalar and spinor wavefunctions.

This will require us to use an orthogonality property that is derived from a 
 conserved current. Let me show this  by 
 defining a generalized conserved current density that involves any two solutions 
 of the Klein-Gordon equation.
 \begin{equation}
	j^{\mu}_{1,2} (x) = i \psi^{*}_{1}(x)\;\stackrel{\leftrightarrow} 
	{\partial^{\mu}} \psi_{2}(x) =  i \psi^{*}_{1}(x)\;[\stackrel{\rightarrow} 
	{\partial^{\mu}} - \stackrel{\leftarrow}{\partial^{\mu}}] 
	\psi_{2}(x) ; \;\;\;
	\partial_{\mu}j^{\mu}_{1,2}(x) 
	= 0.
\end{equation}
Integrating this local conservation law over a 3-dimensional volume 
that contains all of the solutions, we have
\begin{equation}
\frac{d}{dt} \int d^{3}x\; j^{0}_{1,2} (\textbf{x},t) = 0.
\end{equation}
Let me take two plane wave solutions and put them into this integral. 
The integral over $d^{3}x$ requires that the two 3-momenta, 
$\textbf{p}_{1}$, $\textbf{p}_{2}$ be equal. Then we have only two 
possibilities : either $E_{1}= E_{2}$ or $E_{1} = -E_{2}$. But, if it 
is the second case, then the operator 
$\stackrel{\leftrightarrow}{\partial_{t}}$ gives us a zero result. 
This is the orthogonality of positive and negative energy solutions 
of the Klein-Gordon equation.

That is easy for ordinary particles. What about tachyons? We need to 
reconsider how we choose a three-dimensional surface over which to 
integrate $\partial_{\mu}\;j^{\mu}_{1,2}(x) = 0$. We can choose an 
integral over all time and over a two-dimensional space such that all 
of the tachyonic matter in those wavefunctions will pass through that 
surface over time. The integral over all time reqires that the two 
energies be equal. Suppose that the spacial surface is $z=0$: then 
the integral over x and y will require that $p_{x}= p'_{x}$ and 
$p_{y} = p'_{y}$. This leaves us with the two possibilities 
$p_{z} = \pm p'_{z}$. But we see that $j^{z}$ invlves the operator 
$\stackrel{\leftrightarrow}{\partial_{z}}$; and this will provide us 
with the complete orthogonality result.

Now we look at the Dirac equation. Plane wave solutions will  
lead us to the same mass-shell picture; but the orthogonality 
relation comes about differently. For ordinary particles we are 
looking at $j^{0} = \psi^{\dagger}_{1}\psi_{2}$. Looking at the 
spinors shown in (\ref{b1}) we see that they are orthogonal if 
$\textbf{p}_{1} = \textbf{p}_{2}$ and $h_{1}=h_{2}$ but $E_{1} = 
-E_{2}$.

Finally we turn to the Dirac tachyons. The space-time integrals give 
us, as they did with the Klein-Gordon wavefunctions, the possibility of $p_{z} 
= -p'_{z}$. We are then looking at $j^{z} = 
\psi^{\dagger}_{1}\;\sigma_{z}\psi_{2}$.  If $h_{1} = -h_{2}$, we get 
orthogonality from the inner product of the Dirac spinors (\ref{b1}):
\begin{equation}
(E+im)^{*}(E+im) + h_{1}h_{2}p^{2} = 0.
\end{equation}
If we have $h_{1} = h_{2}$, then we have to look at the 2-component 
spinors $|\textbf{p,h}>$.  Here is the relevant theorem.
\begin{equation}
<\textbf{p}, h| \sigma \cdot (\textbf{p} - \textbf{p}')
| \textbf{p}', h>  = <|2p_{z} \sigma_{z} |> = (h-h) p\;<|>=0,
\end{equation}
and that completes the orthogonality proof for $j^{z}$.

\section{Causal Commutators for Ordinary Fields}%7%

For a complex scalar field we have the Klein-Gordon wave equation,
\begin{equation}
[\partial_{t}^{2} - \nabla^{2} + m^{2}] \phi(x) = 0;
\end{equation}
and this leads to the general expansion,
\begin{equation}
\phi(x) = \int\; \frac{d^{3}k}{\sqrt{\omega(2\pi)^{3}}} 
\;\sum_{\epsilon}\;e^{i\epsilon (\textbf{k}\cdot\textbf{x} - \omega 
t)}\; a_{\epsilon} (\textbf{k}), \;\;\;\;\; \omega = 
\sqrt{k^{2}+m^{2}}, \;\;\;\;\; \epsilon = \pm 1.
\end{equation}
The expansion coefficients $a_{\epsilon}(\textbf{k})$ are operators 
that obey some exchange (commutator or anticommutator) relations,
\begin{equation}
[a_{\epsilon}(\textbf{k})\;,\; a^{\dagger}_{\epsilon '}(\textbf{k} ')]_{\pm} = \delta^{3}
(\textbf{k} - \textbf{k}')\;\delta_{\epsilon, \epsilon'}\; ?.
\end{equation}
The unknown quantity $?$ is dimensionless. If we choose to use the 
anticommutator (plus sign on the brackets), then $?$ must be 
positive, for example +1. If we choose the commutator (minus sign on 
the brackets), then $?$ could be of any sign, for example $? = 
\epsilon$ is a posible choice.

Now we calculate the bracket of the field and its conjugate at two 
different space-time points.
\begin{equation}
[\phi(x),\; \phi^{\dagger}(x')]_{\pm} =  \int\; 
\frac{d^{3}k}{\omega(2\pi)^{3}} 
\;\sum_{\epsilon}\;e^{i\epsilon\{ 
\textbf{k}\cdot(\textbf{x}-\textbf{x}') - \omega (t-t')\}}\; ?.
\end{equation}

If we choose $? = +1$, then this integral and sum reduces to
\begin{equation}
- \frac{1}{\pi^{2}\;r}\;\frac{d}{dr}\; \int_{0}^{\infty}\; dk\; cos 
(kr)\; \frac{cos(\omega(t-t'))}{\omega}, \;\;\;\;\; r = |
\textbf{x} - \textbf{x}'|.
\end{equation}

Alternatively, if we choose $? = \epsilon$, we find,
\begin{equation}
 i \frac{1}{\pi^{2}\;r}\;\frac{d}{dr}\; \int_{0}^{\infty}\; dk\; cos 
(kr)\; \frac{sin(\omega(t-t'))}{\omega}.
\end{equation}

This second expresion is causal: the integral vanishes if $r > 
|t-t'|$. This important result may be seen as follows. First extend 
the k-integral to $- \infty$, since the integrand is an even function 
of $k$. Then note that the integrand is an analytic function of 
$\omega^{2}$; thus it is an analytic function of $k$; and we can 
separate $cos (kr) = (e^{ikr}+ e^{-ikr})/2$ and move each contour 
integral into the far upper or lower half-plane. If $r > |t-t'|$, then we 
get zero. 

The previous integral, from $? = +1$, is not an even function of 
$\omega$ and thus it will not give this causal result.

Next, we follow the same line of analysis for the ordinary Dirac equation.
\begin{equation}
[i \frac{\partial}{\partial t} + i \alpha \cdot \nabla - \beta 
m]\;\psi(x) =D\; \psi(x) = 0.
\end{equation}

\begin{eqnarray}
\psi(x) = \int\; \frac{d^{3}k}{(2\pi)^{3/2}}\; \sum_{\epsilon}\; e^{i\epsilon 
(\textbf{k}\cdot\textbf{x} - \omega t)}\;\sum_{h}\; u_{\epsilon, 
h}(\textbf{k})\; b_{\epsilon, h}(\textbf{k}), \\ 
u_{\epsilon, h}(\textbf{k}) = \frac{1}{\sqrt{2\omega(\omega + \epsilon 
m)}}\left( \begin{array}{c} \epsilon \omega +m \\ \epsilon h k  
\end{array}\right) \;|\textbf{k},h>, 
\end{eqnarray}
where $|\textbf{k}, h>$ is an eigenfunction of the 2-component Pauli 
spin matrix dotted into the direction of the momentum vector $\textbf{k}$, with 
eigenvalue (helicity) $h = \pm 1$.

We postulate exchange relations for the expansion coefficients 
(operators),
\begin{equation}
[b_{\epsilon, h}(\textbf{k})\;,\; b^{\dagger}_{\epsilon ', h'}(\textbf{k} ')]_{\pm} = \delta^{3}
(\textbf{k} - \textbf{k}')\;\delta_{\epsilon, \epsilon'}\; 
\delta_{h,h'}\;?,
\end{equation}
and go on to caculate the commutator or anti-commutator of the fields,
\begin{equation}
[\psi(x),\; \psi^{\dagger}(x')]_{\pm} =  \int\; 
\frac{d^{3}k}{(2\pi)^{3}} 
\;\sum_{\epsilon}\;e^{i\epsilon\{ 
(\textbf{k}\cdot(\textbf{x}-\textbf{x}') - \omega (t-t')\}}\; 
\frac{1}{2}\;\sum_{h}u_{\epsilon,h}(\textbf{k})\; u^{\dagger}_{\epsilon, 
h}(\textbf{k})\;?.
\end{equation}
That outer product of $u $ and $u^{\dagger}$ becomes 
\begin{equation}
\frac{1}{\omega} \left( \begin{array}{cc} \omega + \epsilon m & hk 
\\ hk & \omega - \epsilon m \end{array} \right)\; |h> <h|.
\end{equation}
If we choose $? = +1$, we find that this is proportional to
\begin{equation}
\tilde{D} \int_{0}^{\infty} k^{2}\; dk\; \frac{sin(kr)}{kr}\; 
\frac{sin(\omega (t-t'))}{2\omega},
\end{equation}
where we have introduced the adjoint Dirac differential operator 
$\tilde{D}$. The product 
$ D\;\tilde{D}$ is equal to the Klein-Gordon operator.

This last expression is causal, just like the one earlier for a 
scalar field using $? = \epsilon$. However, if we had chosen $? = 
\epsilon$ here 
, our result for the commutator of the Dirac fields would 
not be causal.  One may consider this discussion a (weak and 
incomplete) proof of the 
familiar connection between spin and statistics.

Now we carry out the same study for tachyon fields.

\section{Causal Commutators for Tachyon Fields}%8%

In the Klein-Gordon differential equation we change $m^{2}$ to 
$-m^{2}$ and in the Dirac equation we change $m$ to $im$. We also 
know that the mass shell is quite different and so our expansion of the 
wavefunctions uses somewhat different variables.

For the complex scalar tachyon,
\begin{equation}
\phi(x) = \int_{-\infty}^{\infty}\; d 
\omega\;\frac{\sqrt{k}}{(2\pi)^{3/2}}\; \int d^{2}\hat{k}\;
e^{i(k \hat{k}\cdot \textbf{x} - \omega t)}\; a(\omega, \hat{k}), 
\;\;\;\;\; k = + \sqrt{\omega^{2}+ m^{2}},
\end{equation}
where the three-vector $\textbf{k}$ is now decomposed into 
its magnitude $k$ and its direction $\hat{k}$. We postulate the 
exchange relations,
\begin{equation}
[a(\omega, \hat{k}),\; a^{\dagger}(\omega', \hat{k}')]_{\pm} = 
\delta(\omega - \omega ')\;\delta^{2}(\hat{k} - \hat{k}')\; ?,
\end{equation}
and this is written so that the as yet unknown $?$ is dimensionless.

Now the exchange relation for the fields is,
\begin{equation}
[\phi(x),\; \phi^{\dagger}(x')]_{\pm} = \int_{-\infty}^{\infty}\; 
\frac{d 
\omega}{(2\pi)^{3}} \; e^{-i\omega (t-t')}\; \int d^{2}\hat{k}\; 
e^{ik\hat{k}\cdot(\textbf{x} - \textbf{x}')}\; k\; ?.
\end{equation}
If we try $? = 1$, the result is,
\begin{equation}
 \int_{-\infty}^{\infty}\; d \omega \; e^{-i\omega (t-t')}\;\frac{
 4 \pi}{(2\pi)^{3}} \;
 \frac{sin(kr)}{r}.
 \end{equation}
 This is not causal in any sense.
 
 An alternative choice is $? = \eta\cdot \hat{k}$, where 
 $\eta$ is a vector orthogonal to some chosen surface that 
 is a reference plane for our quantization efforts. This concept was 
 introduced in my earlier (1982) paper on tachyon particles and 
 fields.  If we revert to ordinary (slower then light) fields, then 
 the 4-vector $\eta$ could be taken as $(1,0,0,0)$; and the term 
 $\eta \cdot \hat{k}$ is equivalent to the factor $\epsilon$ used in the 
 previous section.  For tachyons, we might take $\eta = (0,0,0,1)$ or 
 another oriented spacelike 4-vector.
 
 Setting $? = \eta \cdot \hat{k}$, we find for the scalar tachyon 
 fields,
 \begin{equation}
 [\phi(x),\; \phi^{\dagger}(x')]_{-} =-i\frac{ 4 \pi}{(2\pi)^{3}}\; \eta \cdot \nabla \; 
 \int_{-\infty}^{\infty}\; d \omega \; e^{-i \omega (t-t')}\; 
 \frac{\sin(kr)}{kr}.
 \end{equation}
 This is causal for tachyons. This integral will vanish if $|t-t'| > 
 r$.
 
 In my earlier paper I postulated the scalar field commutator to have 
 the factor $? = \eta\cdot k /|\eta\cdot k|$, which is different from what I 
 have used here. That may also lead to a formally causal result but 
 the result found above is much nicer.
 
 Now, we turn to the Dirac tachyon field. 
 
 \begin{equation}
	\psi(x) = \int_{-\infty}^{\infty} \frac{d \omega}{\sqrt{2}(2\pi)^{3/2}}\;  e^{-i \omega t} \; 
	\int d^{2} \hat{k}\; e^{ik \hat{k}\cdot \textbf{x}}\; 
	 \sum_{h}\; \left( \begin{array}{c} \omega 
	+ im \\ hk \end{array} \right) |\hat{k},h> b_{h}(\omega, \hat{k}).
	\end{equation}
	
	Then,
\begin{equation}
[b_{h}(\omega, \hat{k}),\; b^{\dagger}_{h'}(\omega ', \hat{k}')]_{\pm} 
= \delta(\omega - \omega')\;\delta^{2}(\hat{k} - \hat{k}') \; 
\delta_{h,h'} \; ?.
\end{equation}

For $? = 1$, we find
\begin{equation}
[\psi(x),\; \psi^{\dagger}(x')]_{+} = \frac{1}{2(2\pi)^{3}}\tilde {D}\; \gamma_{5}\; 
(\sigma \cdot \nabla/i)\;
\int_{-\infty}^{\infty}\; d \omega\; e^{-i \omega (t-t')}\; \int 
d^{2}\hat{k}\; e^{ik \hat{k}\cdot(\textbf{x} - \textbf{x}')}\;  ,
\end{equation}
and this integral gives the nice tachyon causal result seen above for 
the scalar field. So it looks like we 
can use the anticommutation rules for Dirac tachyons.

In Appendix A we look at this work in spherical coordinates.

\section {Interpreting the number operators}%9%

Combining the formulas from the several previous sections, we 
calculate charge and 4-momentum for the fields in various cases.

For ordinary scalar fields,
\begin{equation}
Q = \int d^{3} x\; j^{0} =  \int d^{3}k\sum_{\epsilon} \label{g1}
a^{\dagger}_{\epsilon}(\textbf{k})\;a_{\epsilon}(\textbf{k})\; \epsilon.
\end{equation}
For $\epsilon = +1$, we interpret each $a^{\dagger}\; a$ as a number 
operator for particles, with $a$ annihilating the vacuum state $|0>$; and for 
$\epsilon = -1$, we interchange the order of operators and take 
$a^{\dagger}$ as the annihilation operator for antiparticles. The 
minus sign between these two number operators says that particle and 
antiparticle have opposite charge. And we learn to ignore the 
infinite constant left over from the commutator.  This is all 
standard business.

For ordinary Dirac fields, we get this result for the charge,
\begin{equation}
Q = \int d^{3}x\;j^{0} = \int d^{3}k\; \sum_{\epsilon, h}\;
b^{\dagger}_{\epsilon, h}(\textbf{k})\;b_{\epsilon, h}(\textbf{k});\label{g2}
\end{equation}
and this for the energy-momentum, 
\begin{equation}
P^{\nu} = \int d^{3}x\; T^{0, \nu} =  \int d^{3} k 
\sum_{\epsilon, h} \; \epsilon \;b^{\dagger}_{\epsilon, h}(\textbf{k})\;
b_{\epsilon,h}(\textbf{k})  (\omega, \textbf{k}).
\end{equation}
Using the anticommutation rule for changing the order of $b$ and 
$b^{\dagger}$ for $\epsilon = -1$, gives us opposite sign of the 
charge for particles and antiparticles; this also gives us positive
energy for 
each particle and also for each antiparticle. Again, zero point 
charge and energy are thrown away.

Now we look at the tachyons.  First, for the scalar field,
\begin{equation}
Q_{\eta} = \int dt\; d^{2}x_{\perp}\; \eta \cdot j = 
\int_{\-\infty}^{\infty} d\omega \; \int d^{2}
\hat{k}\; \eta \
\cdot \hat{k}\; a^{\dagger}(\omega, \hat{k})\; a(\omega, \hat{k});
\end{equation}
and for the Dirac tachyon field,
\begin{equation}
Q_{\eta} = \int_{\-\infty}^{\infty} d\omega \; \int d^{2}
\hat{k}\; \eta \
\cdot \hat{k}\; \sum_{h}\;h\; b_{h}^{\dagger}(\omega, \hat{k})\;
b_{h}(\omega, \hat{k}).\label{g5}
\end{equation}
We look for interpretation of a number operator, like $a^{\dagger}a$,
which will be read as the number of particles flowing through the 
surface with normal $\eta$. (Carefully stated, this is the number of particles 
per unit interval of frequency/energy per unit of solid angle in direction 
of momentum.) For the scalar field, this is just a positive operator.  
For the Dirac field, we see the factor $h$ appears and so we 
anti-commute the operators $b$ and $b^{\dagger}$ for $h=-1$ and then 
we have positive operators with $b^{\dagger}$ identified as the 
destruction operator for the antiparticle.  This is what we had 
anticipated in Section 4.

Finally, we look at the energy-momentum.
\begin{equation}
P_{\eta}^{\nu} = \int dt d^{2}x_{\perp}\;\eta_{\mu}T^{\mu, \nu} = \int_{\-\infty}^{\infty} d\omega \; \int d^{2}
\hat{k}\; \eta \
\cdot \hat{k}\; \sum_{h}\;h\; b_{h}^{\dagger}(\omega, \hat{k})\;
b_{h}(\omega, \hat{k})\;(\omega, \textbf{k}).
\end{equation}	
this appears to give a similarly decent interpretation. If we look at 
$\eta \cdot P$, this becomes a positive quantity - after that 
rearrangement of the b's for h=-1 -  for both particles and 
antiparticles. This is analogous to the positiveness of $P^{0}$ for 
ordinary particles.
	
\section {Lorentz invariance}%10%

Let's look at a couple of ``causal'' integrals found in the previous 
section.

\begin{eqnarray}
\int_{0}^{\infty}\; dk\; cos (kr) \frac{sin(\omega (t-t'))}{\omega} = 
\frac{\pi}{2}\; J_{0} (mT), \;\;\;\;\; T^{2} = (t-t')^{2} - r^{2} > 0; \\
\int_{-\infty}^{\infty} \; d \omega \; e^{-i\omega (t-t')}\; 
\frac{sin(kr)}{k} = \pi\; J_{0}(mS), \;\;\;\;\; S^{2} = r^{2} - 
(t-t')^{2} > 0,
\end{eqnarray}
where $\omega = \sqrt{k^{2}+ m^{2}}$ in the first integral and $ k = 
\sqrt{\omega^{2} + m^{2}}$ in the second integral. The expressions on 
the right-hand side of these equations depend on Lorentz-invariant 
forms involving the space-time coordinate intervals.

Next, some may question whether the use of the reference surface, 
represented by the 4-vector $\eta$, violates Lorentz invariance. 
Formally, any change of coordinate systems should change that vector 
as well; therefore such forms as $\eta \cdot \partial$ are invariant. 
Also, it is worth noting that the ``canonical'' rules for quantizing 
an ordinary field theory involve a particular choice of a reference 
plane, usually $t=0$, in which one introduces operators.

The most challenging question about Lorentz invariance of the theory 
presented here has to do with the identification of helicity as 
marking the distinction between particles and antiparticles for the 
Tachyon Dirac field theory. Is this a Lorentz invariant identification?   It is common to consider the following covariant 4-vector,
\begin{equation}
S_\mu = \frac{i}{2}\;\epsilon_{\mu \nu \kappa \lambda}\;p^\nu\; \gamma^\kappa\; \gamma^\lambda = 
(\mathbf{\sigma} \cdot \textbf{p} ,\; i \textbf{p} \times \mathbf{\alpha} - E \mathbf{\sigma}).
\end{equation}
If we take the expectation value of this operator in one of the plane-wave states described earlier we find,
\begin{equation}
<S_\mu > = <1> \;(hp,\; -hE\hat{\textbf{p}});
\end{equation}
and it is tempting to say that this expression is a timelike 4-vector (for tachyons with $p^2 - E^2 = + m^2$) and therefore the sign of the time component will not change under a Lorentz transformation. 
However, if we just start with a given eigenfunction of helicity and apply a general Lorentz transformation, then some states of the opposite helicity will appear.  Is this a fatal 
flaw of the theory?  I am unsure.  One might consider that the "conservation of helicity" is a principle applicable to any free-particle tachyon state, as somewhat analogous to the conservation of electric charge for ordinary particles. Perhaps this should be regarded as an unresolved question about what physical properties such particles, if they exist, might have. 

In earlier work \cite{CS2} it was pointed out that the distinction between 
``absorption'' and ``creation'' of a tachyon was not a Lorentz 
invariant concept - although it is so for ordinary (slower then 
light) particles. Perhaps a similar argument applies to the 
disinction beteen particle and antiparticle (absent such a conserved 
quantity like electric charge). Such considerations appear in familiar 
theoretical studies \cite{BK} of neutrinos in the Majorana representation.

In the limit of Dirac theory with zero mass, that is to say looking 
at the Weyl 2-component equation, helicity is a Lorentz invariant.  
So, one might predict, if the above theory is accepted, that there 
may be mixing of particle and antiparticle in the quantitative amount 
of order $m/p$. How might one detect such a phenomenon, for example, 
if neutrinos were tachyons, with mass less than 1 eV?

\section{ Dealing with der Gruppenpest}%11%

There is a familiar line for dismissing tachyons in quantum theory that goes like this: "The Little Group for tachyons is O(2,1) and this has no unitary representations aside from the one-dimensional."

Let's examine this problem.  The group theory approach for finding irreducible representations of the Lorentz group (initiated by Wigner many years ago) starts by identifying a subgroup of Lorentz transformations - called the Little Group - that leave unchanged the four-momentum of a single particle state as seen in some particular frame of reference. For a normal  (slower than light) particle this is the rest frame (3-momentum equal to zero) and the Little Group is O(3), any rotation in 3-space.  

For a tachyon, the selected frame of reference sees the particle with energy equal zero and momentum oriented in some direction, call it the z-axis. That 4-vector is invariant under the following transformations: rotation about the z-axis and boost in the x-y plane. This Little Group is generated by the three operators $J_z, K_x, K_y$ which obey the following Lie algebra,
\begin{equation}
[J_z, K_x] = i K_y, \;\;\; [K_y, J_z] = i K_x, \;\;\; [K_x, K_y] = -i J_z,
\end{equation}
and it is the minus sign in that last commutator that says this is the group O(2,1) and not O(3). The invariant form is $J_z^2 - K_x^2 - K_y^2 $.
Now we can easily write down a 2-dimensional complex representation of this Lie algebra, as follows,
\begin{equation}
J_z = \frac{1}{2} \sigma_3, \;\;\;\;\; K_x = \frac{i}{2} \sigma_1, \;\;\;\;\; K_y = \frac{i}{2} \sigma_2,
\end{equation}
using the standard Pauli spin matrices. We note that two of those operators are not Hermitian; so, when we form a general operator for the Lie Group,
\begin{equation}
U = e^{i \alpha J_z + i \beta K_x + i \gamma K_y},
\end{equation}
where $\alpha, \beta, \gamma$ are arbitrary real parameters, this is not a unitary matrix.

We note the following identity, involving the Hermitian adjoint and the inverse of this matrix $U$.
\begin{equation}
U^\dagger = \sigma_3\; U^{-1}\; \sigma_3.
\end{equation}
This leads us to construct a new rule for inner products in the space of this representation, one that uses $\sigma_3$ as a metric and leads to the following invariance property:
\begin{equation}
<\psi ' | \sigma_3 | \psi> \longrightarrow <U\psi '| \sigma_3 |U \psi> = 
<\psi '| U^\dagger \sigma_3 U |\psi> = <\psi '| \sigma_3 |\psi>.
\end{equation}

Is there any harm caused by introducing a metric into the Hilbert space of a quantum theory? There may be objections in general but let us see what is actually involved in the particular situation being studied here, namely, the state of a single free particle that happens to be a tachyon with spin. This operator $\sigma_3$ has eigenvalues $\pm 1$; and we recognize this as being exactly the helicity described in our earlier study of the Dirac equation for a tachyon. (Remember that the z-direction is the direction of the particle's momentum in the selected frame of reference.)

So, here is where we stand. We can handle the Little Group for spin 1/2 tachyons if we use the helicity operator as a metric in the Hilbert space for a single particle or a single antiparticle. This leads us back to the previous discussion about helicity being the quantity that defines particles vs antiparticles among solutions of the Dirac equation for tachyons.

(I wonder what would happen if we introduced a similar nontrivial metric  - say $\epsilon$,  the sign of the energy - in the group theory analysis for ordinary particles. When we looked at the conserved 4-current for ordinary solutions of either the Klein-Gordon or the Dirac equations we found such a plus-or-minus sign in the inner product that is identified as the charge.)

\section{Conclusion and Prospectus}%12%

We have achieved substantial success in formulating a consistent quantum theory for free tachyon fields. This was achieved by starting with a particular basis in momentum space, $(\omega, \hat{k})$, 
and then selecting commutation or anticommutation rules that led to explicit tachyon causality.

What physical interpretation do we give to the conserved current density?  For ordinary particles, scalar or spinor, we use $j_\mu (x)$ to couple to the electromagnetic field; and so $Q = \int d^3 x \; j_0$ is identified with the electric charge of the particle. In the quantized expressions Eqs. (\ref{g1}) and (\ref{g2}) we see that this has one sign for the particle and the opposite sign for the antiparticle. This is familiar for ordinary particles. What about the tachyons fields that we have now quantized? Looking at Eq. (\ref{g5}) for the Dirac tachyon, we see the same sign for $Q_\eta$ for both the particle and the antiparticle. This suggests that this field does not couple to the electromagnetic field.  It will be, of course, a matter of experimental fact whether such a spinor tachyon system carries electric charge - if such a thing actually exists; but the mathematical work we present here does make this suggestion. Thus neutrinos, which carry no net electric charge, are an enticing candidate for tachyonicity.

Earlier theoretical studies of a possible charged tachyon had quite a time making sense out of Cherenkov radiation that would be expected from such a particle. If the neutrino is seriously considered as a candidate, one might ask whether there is some internal charge structure that might lead to such radiation effects. A recent paper \cite{CG}  provides some formulas that can be used to estimate this effect; and, if one assumes a mass around 1 eV for a tachyon neutrino,  it is small in the extreme, even over cosmic time. Another paper \cite{CS3} provides a calculation of the gravitational Cherenkov radiation from such a tachyon and, again, finds that it is small in the extreme even over cosmic times.

The KATRIN (KArlsruhe TRitium Neutrino) experiment, now in progress, will be very relevant to the question of whether the neutrino may be a tachyon.

What remains to be studied on the theoretical side?  Introduce interactions with other fields; formulate an appropriate extension of the S-matrix; see how tachyons might be incorporated into the Standard Model of elementary particles.

\newpage
\textbf{Acknowledgment}

I thank K. Bardakci for some helpful discussion.

\vskip 0.5cm
\setcounter{equation}{0}
\def\theequation{A.\arabic{equation}}
\boldmath
\noindent{\bf Appendix A: Scalar tachyon field  quantized on a sphere}
\unboldmath
\vskip 0.5cm

Let's try the expansion of a complex scalar field for tachyons in 
terms of familiar functions for spherical geometry.
\begin{eqnarray}
\phi(x) = \int d\omega e^{-i\omega t}\sum_{l,m}\; Y_{lm}(\hat{x})\;
[a_{l,m}(\omega) h_{l}^{(1)}(kr) + b^{\dagger}_{l,m}(\omega) h_{l}^{(2)}(kr)], 
 \label{A1} \\
\;\;[a_{l,m}(\omega),a^{\dagger}_{l',m'}(\omega')] = 
[b_{l,m}(\omega),b^{\dagger}_{l',m'}(\omega')] = c k \delta(\omega - 
\omega ') \delta_{l,l'}\delta_{m,m'},
\end{eqnarray}
where all other commutators vanish and $k=\sqrt{\omega^{2}+m^{2}}$. 
We have separated the terms representing outgoing waves, 
$h_{l}^{(1)} \sim e^{ikr}/kr$,  from incoming waves 
$h_{l}^{(2)} = h_{l}^{(1)\; *}$.

We calculate the commutator of the field operators and get:
\begin{eqnarray}
[\phi(x),\;\phi^{\dagger}(x')] = \int d\omega e^{-i\omega (t-t')}\;
\sum_{l,m}\;Y_{lm}(\hat{x})\;Y^{*}_{lm}(\hat{x}') \;\\ \nonumber 
ck[h_{l}^{(1)}(kr) h_{l}^{(2)}(kr')- 
h_{l}^{(2)}(kr)h_{l}^{(1)}(kr')].\label{A3}
\end{eqnarray}
If we take the 
derivative with respect to $r'$ and then set $r'=r$, we get the 
analog of the usual equal time commutator of conjugate fields.
\begin{equation}
	[\phi(r,\hat{x},t),\frac{\partial}{\partial r} \phi^{\dagger}( 
	r,\hat{x}',t')] = -i \delta(t-t')\;\delta^{2}(\hat{x} - 
	\hat{x}')/ r^{2},
\end{equation}
with the choice $c=1/(4 \pi) $.  

Moreover, one can show  that the 
general commutator Eq. (\ref{A3}) will vanish if $|t-t'| > |r-r'|$, which seems to 
be an overstatement of the idea that tachyons always 
travel faster than light. However, if we look at large $r$ and $r'$ 
(very large compared to 
1/m), then we might substitute the asymptotic values for the spherical 
Hankel functions in that equation and get the following formula.
\begin{equation}
[\phi(r,\hat{x},t), \phi^{\dagger}( r',\hat{x}',t')] \approx  i \;\delta^{2}(\hat{x} - 
	\hat{x}')) 
	\int_{-\infty}^{\infty} d\omega\;e^{-i\omega (t-t')}\; 
	\frac{sin k(r-r')}{2 \pi krr'}.
	\end{equation}
This shows an unexpected result that 
the general commutator vanishes unless $\hat{x} = \hat{x}'$; and in that context 
this tachyonic ``causal'' restriction  makes better sense.

There appears to be some delicate math here and we can suspect that 
working under the infinite sum over $l$ we have made some careless 
assumptions about convergence.

Consider, for example, this Green's function for the tachyonic Klein- 
Gordon equation.
\begin{equation}
G_{0}(x,x') = \int_{-\infty}^{\infty}\;d \omega\;e^{-i\omega (t-t')}\; 
\frac{cos kR}{R}, \;\;\;\;\;R = |\textbf{x} - \textbf{x}'|.\label{B4}
\end{equation}
This will vanish if $|t-t'| >R$, as one sees using arguments given in 
Section 7. However, when this is expanded in spherical functions, 
 analogous to (\ref{A3}), the relevant 
functions are $k\;j_{l}(kr_{<})\;n_{l}(kr_{>})$, and this  would 
lead us to  the wrong limit: each term in the sum over $l$ vanishes if 
$|t-t'| > |r+r'|$.

In summary, we have seen that expanding the field in spherical functions gives a less-than-perfect result for causality of the tachyon field operators; however,  we did find a perfect analog for the "equal-time" commutation rule.  By contrast, the earlier work, in Section 8, gave a perfect causal result for the tachyon field operators; but that did not give us the analog of "equal-time" commutators.

\end{document}